\documentclass[twoside]{dis04}
\def\runauthor{T.\ Gleisberg et al.}
\def\shorttitle{New tools for automatic cross section calculation}
\begin{document}

\title{New tools for automatic cross section calculation}

\author{T.\ Gleisberg, S.\ H{\"o}che, F.\ Krauss, A.\ Sch{\"a}licke,
        S.\ Schumann, \\ J.\ Winter, G.\ Soff}

\address{Institute for Theoretical Physics\\
TU Dresden, {\bf D}-01062 Dresden, Germany\\
E-mail: krauss@theory.phy.tu-dresden.de}

\maketitle

\abstracts{In this contribution the matrix element generator {\tt AMEGIC++}
           will be presented. It automatically generates Feynman diagrams,
           helicity amplitudes, and suitable phase space mappings for 
           processes involving multi-particle final states within the Standard
           Model and some of its popular extensions.}

Due to the rising energy frontier in high-energy collider experiments, potential 
signals and their backgrounds for interesting or new phenomena involve an increasing 
number of final state particles and their correlations. To describe the production 
process of a number of particles, in a quantum mechanically correct treatment, at 
least at leading order, the corresponding amplitudes have to be constructed. This 
usually results in a potentially very large number of terms, such that their automated 
construction and evaluation becomes mandatory. Apart from handling the 
sheer number of amplitudes, which grows factorially or 
worse with the number of final state particles, the integration over the multidimensional 
phase space of the final state represents a formidable task. In the past years various
solutions, implemented as different codes, to surmount these problems, have been found.
Codes, which incorporate the full Standard Model, are:

\noindent
{\tt CompHEP} \cite{Boos:1994xb}, which relies on the traditional method of constructing 
and summing Feynman diagrams, where completeness relations are used to square the total 
amplitude. The integration is achieved through one phase space mapping selected by the user.
{\tt MadGraph/MadEvent} \cite{Stelzer:1994ta},
which constructs Feynman 
diagrams and employs the method of helicity amplitudes through the {\tt HELAS} library 
\cite{Murayama:1992gi} for their evaluation. The phase space integration is achieved 
through a single-diagram enhanced mapping.
There, each diagram 
gives rise to one parametrisation of phase space, their interplay is steered dynamically in 
order to minimise the overall variance.
{\tt Alpgen} \cite{Mangano:2002ea} uses the (extended) $\alpha$-formalism 
\cite{Caravaglios:1995cd}
to construct the amplitudes.
This formalism is based on the Schwinger-Dyson method to recursively define one-particle 
off-shell Greens functions, which are then numerically evaluated through a specific 
representation of their ingredients. This approach significantly tames the factorial
growth of the number of terms to be calculated with the number of final state particles.
In {\tt Alpgen} a huge number of processes at hadron colliders is implemented ready to
use, the amplitudes are supplemented with suitable, predefined phase space mappings. 
The {\tt HELAC/PHEGAS} package \cite{Kanaki:2000ey}
also employs the Schwinger-Dyson method, however, in a slightly different fashion. 
The phase space mappings are constructed automatically after Feynman diagram-like
topologies underlying the amplitudes have been defined.

\begin{figure}
\begin{center}
\begin{tabular}{cc}
{\includegraphics[width=4cm]{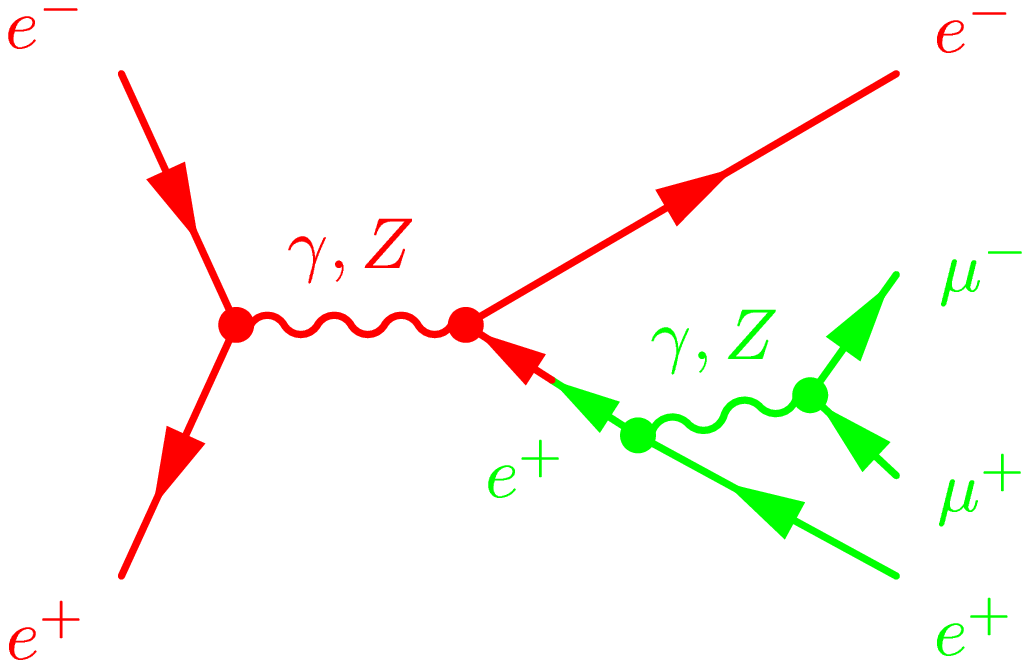}} &
{\includegraphics[width=4cm]{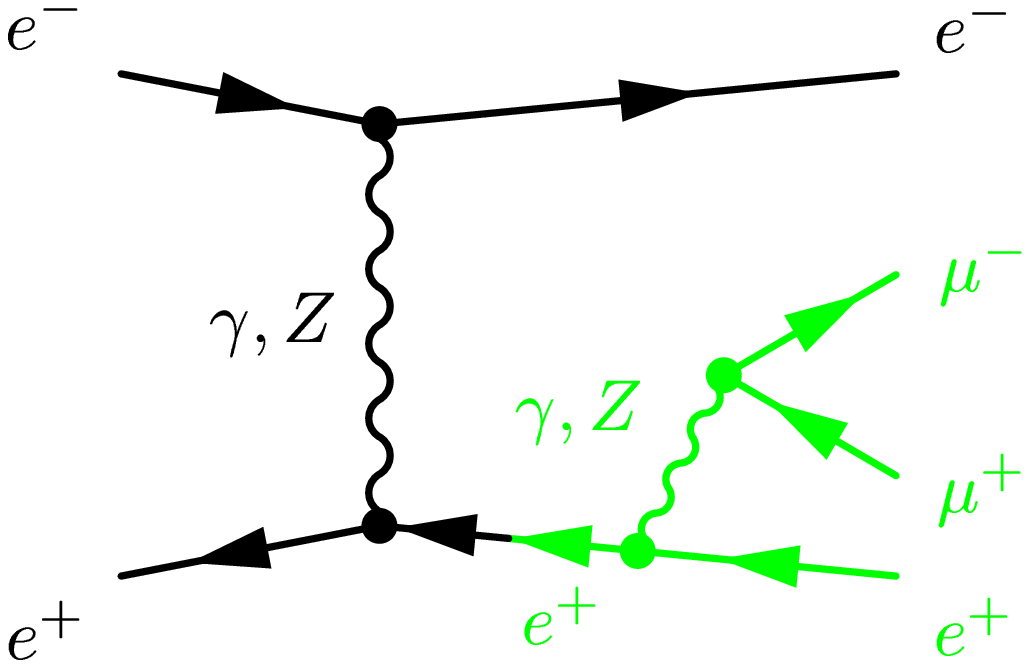}}
\end{tabular}
\caption{\label{SuperampFig}Common parts of the amplitudes (in green) can be factored
         out, reducing the number of complex multiplications, leading to ``super-amplitudes''.} 
\end{center}
\end{figure}

\begin{figure}
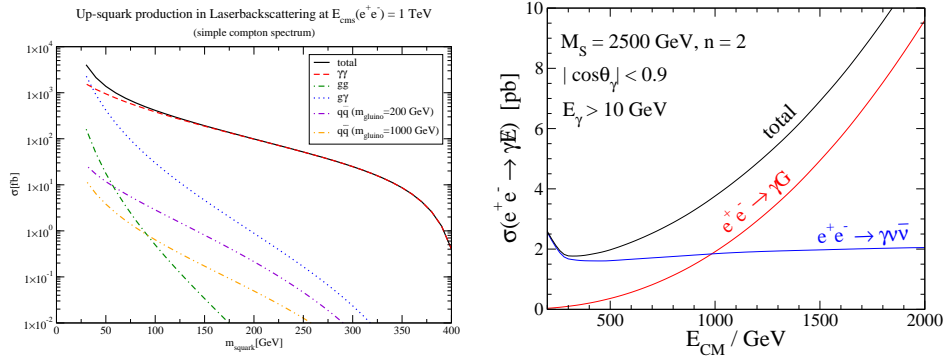

\begin{tabular}{cc}
\begin{minipage}{6cm}
\begin{center}
\includegraphics[width=6cm]{gg_squark_new2.eps}
\end{center}
\end{minipage} &
\begin{minipage}{6cm}
\begin{center}
\includegraphics[width=6cm]{ee_gG.eps}
\end{center}
\end{minipage} 
\end{tabular}
\caption{\label{squarks}Production of squarks at a photon collider (by laser-backscattering 
         off an $e^+e^-$ collider) including the effect of the photons hadronic structure
         (left plot). In the right plot the production cross section for real ADD-gravitons is shown, in dependence of the c.m.-energy of the colliding
         $e^+e^-$-pair.}
\end{figure}

\noindent
In this talk, however, the focus will be on {\tt AMEGIC++} \cite{Krauss:2001iv}
which constructs Feynman diagrams and helicity amplitudes in the fashion of 
\cite{Kleiss:1985yh}
In order to reduce the factorial growth, symbolical 
manipulations are performed, cf. Fig. \ref{SuperampFig}, which significantly limit the number 
of terms to be evaluated, see also Table \ref{SuperampTab}. The helicity method has been 
extended to spin-2 particles allowing {\tt AMEGIC++} to include some models featuring graviton 
resonances \cite{Gleisberg:2003ue}. The phase space mappings are generated automatically from 
the Feynman diagrams. The integration is performed using standard multi-channel methods \cite{Kleiss:qy}. 
\begin{table}
\begin{center}
{\small
\begin{tabular}{|l|r|r|}
\hline\hline
$e^+e^- \to$ & Feynman diagrams& ``Super-amplitudes''\\\hline   
$e^+e^-$ & 4 & 4\\
$e^+e^-\mu^+\,\mu^-$ & 50 & 10\\
$e^+e^-e^+e^-$ & 144 & 9\\
$e^+e^-e^+e^-\mu^+\mu^-$ & 3690 & 261\\
$e^+e^-e^+e^-e^+e^-$ & 13896 & 323\\
\hline\hline
\end{tabular}
}
\caption{\label{SuperampTab}Effect of factoring out common parts of the amplitudes, leading
         to ``super-amplitudes''.} 
\end{center}
\end{table}
{\tt AMEGIC++} has been tested for a plethora of processes. For $e^+e^-$ colliders, there is an
ongoing study in the framework of the ECFA-DESY workshop to validate such tools for six-body
final states; results of such a comparison of {\tt AMEGIC++} with {\tt HELAC/PHEGAS} 
\cite{Gleisberg:2003bi} are presented in the left panel of Table \ref{sixbodiesTab} and in its right panel, relative deviations are displayed. Obviously the errors are distributed statistically, 
hence the results of the two codes are in perfect agreement. 
\begin{table}
\begin{tabular}{cc}
\begin{minipage}{6cm}
\begin{center}
{\tiny
\begin{tabular}{lccc} 
\hline\noalign{\smallskip}      
final state          & QCD & {\tt AMEGIC++} [fb] & {\tt HELAC} [fb]\\
\noalign{\smallskip}\hline\noalign{\smallskip}
$b\bar b u \bar dd \bar u$ 
                     & yes & 49.74(21) & 50.20(13)\\    
                     & no & 49.42(44)  & 50.55(26)\\
$b\bar b u \bar ug g$ 
                     & -- & 9.11(13)  & 8.832(43)\\
$b\bar b g gg g$ 
                     & -- & 24.09(18) & 23.80(17)\\
$b\bar b u \bar d e^- \bar \nu_e$ 
                     & yes & 17.486(66) & 17.492(41)\\
                     & no & 17.366(68)  & 17.353(31)\\
$b\bar b e^+ \nu_e e^- \bar \nu_e$ 
                     & -- & 5.954(55) & 5.963(11)\\
$b\bar b e^+ \nu_e \mu^- \bar \nu_{\mu}$ 
                     & -- & 5.865(24) & 5.868(10)\\
$b\bar b \mu^+ \nu_{\mu} \mu^- \bar \nu_{\mu}$ 
                     & -- & 5.840(30) & 5.839(12)\\
\noalign{\smallskip}\hline
\end{tabular}}
\end{center}
\end{minipage} &
\begin{minipage}{6cm}
\begin{center}
\includegraphics[width=4cm]{sixfermana.eps}
\end{center}
\end{minipage} 
\end{tabular}
\caption{\label{sixbodiesTab} Some cross sections for processes with
         six-body final states in $e^+e^-$ collisions (left panel). 
         In the right figure, relative deviations of all cross sections calculated 
         independently by {\tt AMEGIC++} and {\tt HELAC/PHEGAS} are shown, proving that the 
         results of the two codes coincide.}
\end{table}
In the framework of the MC4LHC workshop \cite{MC4LHC} in a similar fashion, different programs for
the calculation of cross sections in proton-proton collisions have been compared; exemplary
results can be found in Table \ref{hadron1}.
\begin{table}[ht]
\begin{center}
{\small
\begin{tabular}{|c|c|c|c|c|c|c|c|c|}
\hline
\multicolumn{2}{|c|}{ X-sects (pb)} & \multicolumn{6}{c|}{ Number of jets}\\\hline
\multicolumn{2}{|c|}{ $ e^- \bar \nu_e $ + $n$ QCD jets }& 0 & 1 & 2 & 3 & 4  & 5 \\
\hline
\multicolumn{2}{|c|}{Alpgen}   & 3904(6)& 1013(2) & 364(2)& 136(1) & 53.6(6) & 21.6(2) \\
\multicolumn{2}{|c|}{CompHEP}  & 3947.4(3)& 1022.4(5)& 364.4(4)& & & \\
\multicolumn{2}{|c|}{MadEvent} & 3902(5)& 1012(2)& 361(1)& 135.5(3) & 53.6(2) & \\
\multicolumn{2}{|c|}{Amegic++/Sherpa}   & 3908(3) & 1011(2) & 362(1) & 137.5(5) & 54(1)  &\\
\hline
\end{tabular}\\[5mm]
\begin{tabular}{|c|c|c|c|c|c|c|c|}
\hline
\multicolumn{2}{|c|}{ X-sects (pb)} & \multicolumn{5}{c|}{ Number of jets}\\\hline
\multicolumn{2}{|c|}{ $ e^- \bar \nu_e $ + $b\bar b$ }& 0 & 1 & 2 & 3 & 4  \\
\hline
\multicolumn{2}{|c|}{Alpgen}   & 9.34(4)& 9.85(6)& 6.82(6)& 4.18(7)& 2.39(5) \\
\multicolumn{2}{|c|}{CompHEP}  & 9.415(5)& 9.91(2)& & &  \\
\multicolumn{2}{|c|}{MadEvent} & 9.32(3)& 9.74(1)& 6.80(2)&  & \\
\multicolumn{2}{|c|}{Amegic++/Sherpa}   & 9.37(1) & 9.86(2) & 6.87(5) &   &   \\
\hline
\end{tabular}
}
\caption{\label{hadron1}Compilation of results for cross sections of some selected processes at the LHC.}
\end{center}
\end{table}
Furthermore, some results for production cross sections of exotic particles in the framework of 
supersymmetric extensions of the Standard model and of models for extra dimensions are presented 
in Figs \ref{squarks}. In the left plot the cross section for the production of up-type squarks by 
laser-backscattered photons (with corresponding Compton spectrum)
at an $e^+e^-$-collider is depicted, displaying also the contributions due to the hadronic structure 
of the photons. In the right panel of this figure, the production cross section for graviton 
resonances in the ADD model \cite{Arkani-Hamed:1998rs}, also at an $e^+e^-$-collider, is exhibited 
in dependence on the c.m.-energy of the incoming $e^+e^-$-pair.

\noindent
It should be noted that {\tt AMEGIC++} is an integral part of a full-fledged event generator called
{\tt SHERPA} \cite{Gleisberg:2003xi}. The connection to its internal parton shower \cite{Kuhn:2000dk}
is achieved according to the merging prescription of \cite{Catani:2001cc}
rendering it an unique tool. It is anticipated that {\tt SHERPA} will also contain a new version of
the cluster hadronisation, presented in \cite{Winter:2003tt}. Taken together, {\tt AMEGIC++} 
within this framework represents a powerful tool for studying various relevant production
processes within and beyond the Standard Model and for physics simulations at current and future
colliders.

\section*{Acknolwedgments}
The authors gratefully acknowledge financial support by BMBF, DFG, and GSI. F.K.\ wishes to thank
the organisers of DIS2004 for the extremely pleasant and fruitful atmosphere.

\end{document}